\documentstyle[11pt,bezier]{article}

\input epsf
\newcommand{\Ref}[1]{(\ref{#1})}
\title{\bf The Reggeon $\to$ 2 Reggeons $+$ Particle vertex
in the Lipatov effective action formalism}
\author{M.A.Braun and M.I.Vyazovsky\\
Dep. of High Energy physics,
 University of S.Petersburg,\\
198904 S.Petersburg, Russia}
\def\beq{\begin{equation}}
\def\eeq{\end{equation}}
\def\tr{{\rm Tr}\,}
\def\pd{\partial}
\def\ep{\epsilon^*_\perp}

\def\noi{\noindent}
\begin{document}
\maketitle
\medskip
\noi{\bf Abstract.}
The vertex for gluon emission during the splitting of a reggeized gluon into two is
constructed in the framework of Lipatov effective action formalism. Its reduction to
a pure transverse form for the diffractive amplitude
gives the standard Bartels vertex plus an additional
contribution corresponding to the emission from a pointlike splitting vertex. This additional
contribution turns out to be given by a longitudinal integral divergent both in the
ultraviolet and infrared. A certain specific recipe for this part, including the
principal value prescription for the integration, allows to eliminate this
unwanted contribution.

\vspace{0.5cm}

\section{Introduction}
Particle interaction in the Regge kinematics, when energy $\sqrt{s}$ is large and much
greater than the typical transferred momentum $\sqrt{-t}$, in the perturbative QCD is
described by
exchanges of reggeized gluons ('reggeons') accompanied by emission of real gluons
('particles').
In the high-colour approximation this picture reduces to the exchange of BFKL
pomerons
~\cite{BFKL}, which fuse and split with known triple-pomeron vertexes
~\cite{bar,mul,brava}.
Summation of the resulting diagrams in the tree approximation leads to the BK equation for
DIS ~\cite{bal,kov,bra1} or a pair of equations for in- and outgoing pomerons for
nucleus-nucleus
scattering ~\cite{braab}. Contribution from loop diagrams may be studied in the formalism
of an effective field theory ~\cite{braeff} or in the competing Hamiltonian formalism
in terms of gluon variables (see e.g. ~\cite{hamglu} and references therein).

In all these developments the basic blocks have been calculated  directly from
the relevant 4-dimensional Feynman diagrams in the Regge kinematics and converted
into the corresponding expressions in the 2-dimensional transverse space
(momentum or coordinate). This procedure is easily realized for simplest elements,
such as a particle emission during the exchange of  a single reggeized gluon,
 but becomes very cumbersome for more complicated processes. A potentially
powerful formalism of an effective action has been proposed by L.N.Lipatov
~\cite{lip}, in which the longidudinal and transverse variables are separated from
the start and one arrives at a theory of interaction of reggeized gluons and particles
described by independent fields. This formalism, in principle, allows to
automatically calculate all diagrams in the Regge kinematics in a systematic and
self-consistent way. However the resulting vertexes are  4-dimensional and
reduction of them to the 2-dimensional transverse form  still has to be done.

Up to the present several application of this formalism have been done and a number of
interaction vertexes have been calculated. They include vertexes for the transition
of a reggeon into several particles or into a reggeon and several particles
~\cite{ALKC}. In this paper we study a vertex for the transition of a reggeon
into a pair of reggeons and a particle (the RRRP vertex in the terminology of
~\cite{ALKC}) in the effective
action formalism.
The 2-dimensional form of this vertex (the 'Bartels' vertex) is well-known
~\cite{bar,BW} (see also ~\cite{muldoc} for its form in the coordinate space).
The found 4-dimensional vertex  resembles the Bartels vertex, although it
contains a new structure absent in the latter and of course longitudinal
variables.
Unlike the vertexes obtained so far in the effective action formalism,
reduction of the RRRP vertex  to the 2-dimensional form involves a non-trivial
integration in the loop formed by the two reggeons and the target. We discuss this
integration and show that for a diffractive diagram, in which both the
projectile and target are just quarks, a literal integration over the
longitudinal variables is impossible (divergent). The Bartels vertex is obtained
only if certain {\it ad hoc} rules are followed, which reduce to  neglecting all
small longitudinal momenta in the target factor and subsequent
integration over the minus component of the loop momentum according to the
principal value prescription.

\section{Feynman rules from the effective action}
These were presented in ~\cite{ALKC} and we only recapitulate them here for
convenience and to fix our notations. We use the light-cone variables defined
by light-cone unit vectors
$n_{\mu}^{\pm}=(1,0,0,\mp 1)$. So
$a_{\pm}=a_\mu n^{\pm}_{\mu}=a_0\pm a_3$
and $ab=a_\mu b_\mu=(1/2)(a_+b_-+a_-b_+)+a_\perp b_\perp$.
The metric tensor is  $g_{+-}=g_{-+}=1/2$, $g_{11}=g_{22}=-1$
We standardly denote $D_{\mu}=\partial_{\mu}+gV_{\mu}$ where
$V_{\mu}=-iV_{\mu}^aT^a$ is the gluon field and  $T^a$ are the $SU(N)$
generators in the adjoint representation.
The reggeon field  $A_{\pm}=-iA_{\pm}^aT^a$  satisfies the kinematical
condition
\beq\partial_+A_-=\partial_-A_+=0.
\label{e1}\eeq
The field $A_+$ comes from the region with a higher rapidity, his momentum
$q_-$ is small, the field $A_-$ comes from the region with a smaller rapidity,
his $q_+$ is small.
The QCD Lagrangian for the particle (gluon) field $V$ is standard and so are the
Feynman rules.
Parts of effective action describing  the reggeon field $A$ are
presented in ~\cite{lip}, to which we refer as L .
The effective action is (L.210)
\beq{\cal L}_{eff}={\cal L}_{QCD}(V_\mu+A_\mu)+
\tr\Big(({\cal A}_+(V_++A_+)-A_+)\pd^2_{\perp} A_-+
({\cal A}_-(V_-+A_-)-A_-)\pd^2_{\perp} A_+\Big),
\label{e2}\eeq
where
\[ {\cal A}_{\pm}(V_{\pm})=-\frac{1}{g}\pd_{\pm}\frac{1}{D_{\pm}}\pd_{\pm}*1=
\sum_{n=0}(-g)^nV_{\pm}(\pd_\pm^{-1}V_\pm)^n
\]\beq=
V_{\pm}-gV_{\pm}\pd_\pm^{-1}V_\pm+g^2V_{\pm}\pd_\pm^{-1}V_\pm
\pd_\pm^{-1}V_\pm+ -...
\label{e3}\eeq
The shift $V_\mu\to V_\mu+A_\mu$ with $A_\perp=0$ is to exclude direct
transitions  $V\leftrightarrow A$.

The propagator is determined by terms
quadratic in the fields.
In the Feynman gauge ${\cal L}_{QCD}^{quadr}=\tr(-V_\mu\pd^2V_\mu)$,
which after substitution $V_{\pm}\to V_\pm+A_\pm$ leads to
\[
\tr\Big(-(V_\nu+A_\nu)\pd^2(V_\nu+A_\nu)\Big)\]
\beq=
\tr(-V_\nu\pd^2V_\nu)+\tr(-V_+\pd_\perp^2A_--V_-\pd_\perp^2A_+)+
\tr(-A_-\pd_\perp^2A_+).
\label{e4}\eeq
The second term cancels with a similar term in the induced action and the
quadratic terms which remain are
\beq
\tr(-V_\nu\pd^2V_\nu)+\tr(-A_-\pd_\perp^2A_+)=
\frac{1}{2}V_\nu^a\pd^2V_\nu^a+\frac{1}{4}(A_-^a\pd_\perp^2A_+^a+
A_+^a\pd_\perp^2A_-^a).
\label{e5}\eeq
Thus the reggeon propagator in the momentum space is
\beq
<A_+^aA_-^b>=-i\frac{2\delta_{ab}}{q_\perp^2}.
\label{e6}\eeq

The reggeon$\to$ reggeon $+$ particle ('Lipatov') vertex (Fig. 1.1) is
\beq
\frac{gf^{ab_2d}}{2}\Big[ q_\sigma+q_{2\sigma}+
\Big(\frac{q_2^2}{q_+}-q_{2-}\Big)n^+_\sigma+
\Big(\frac{q^2}{q_{2-}}-q_+\Big)n^-_\sigma\Big].
\label{e7}\eeq
Note that in the gauge $V_+=0$ (for $p_A+p_B$ collision it is equivalent to $p_BV=0$)
the polarization vector has the form
\beq
\epsilon_\mu(k)=
\epsilon^{\perp}_\mu(k)-\frac{k\epsilon^{\perp} }{k_+} n_\mu^+.
\label{e8}\eeq
Convolutiong the vertex \Ref{e7} with this polarization vector we find
\beq
 gf^{ab_2d}q_{\perp}^2\Big(\frac{q\epsilon_\perp}{q_\perp^2}-
\frac{k\epsilon_\perp}{k_\perp^2}\Big)
\label{e9}\eeq
This gives the form of the  Lipatov vertex widely used in literature.
\begin{figure}[ht]
\epsfxsize 4in
\centerline{\epsfbox{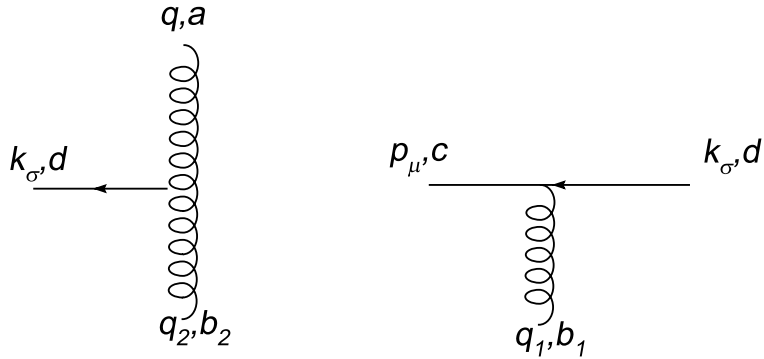}}
\caption{Vertexes reggeon$\to$ reggeon+particle (1) and
particle$\to$reggeon+particle (2)}
\label{Fig1}
\end{figure}

Other ingredients of the technique will be derived in the course of the derivation.
They use parts of the effective action listed below.
The part containing two fields $A_-$ is (L.276)
\[
{\cal L}_{A_-A_-}=-\frac{1}{4}\tr\Big([D_+,A_-]\Big)^2=
-\frac{1}{4}\tr\Big([\partial_+-igV_+^aT^a,-iA_-^bT^b]\Big)^2
\]\beq=
-\frac{1}{4}\tr\Big(\partial_+A_-^b(-iT^b)+gV_+^aA_-^b(-if^{abc}T^c)\Big)^2.
\label{e10}\eeq
The part  containing one field $A_-$ and one $A_+$ is (also
L.276)
\[ {\cal L}_{AAV}=g\tr\Big(
[V_\sigma^{\perp},A_+]\pd_\sigma^\perp A_- +
[V_\sigma^{\perp},A_-]\pd_\sigma^\perp A_+
\]\beq +
\frac{1}{2}[A_-,A_+] \pd_+V_-+  \frac{1}{2}[A_+,A_-] \pd_-V_+\Big)+
g\tr\Big( -[V_+,\pd_+^{-1}A_+]\pd_\perp^2A_-
 -[V_-,\pd_-^{-1}A_-]\pd_\perp^2A_+\Big).
 \label{e11}\eeq
Here the first big bracket corresponds to the standard triple vertex and the
second to the induced one.
The part of the action containing two fields $A_-$, one $A_+$ and $V_\mu$ is
(L.277)
\beq
{\cal L}_{AAAV}=-\frac{g}{2}\tr \Big(-[D_+,A_-][A_-,A_+]+
2\pd_-\frac{1}{D_-}A_-\frac{1}{D_-}A_-\frac{1}{D_-}\pd_-\pd^2_{\perp}A_+\Big).
\label{e12}\eeq
Here
\beq\frac{1}{D_-}=\frac{1}{\pd_-+gV_-}=\pd_-^{-1}\frac{1}{1+g\pd_-^{-1}V_-}=
\pd_-^{-1}\sum_{n=0}(-g\pd_-^{-1}V_-)^n=
\pd_-^{-1}-g\pd_-^{-1}V_-\pd_-^{-1}  +...\label{e13}\eeq
Finally the part of the action containing two gluon and one reggeon field is
(L.275)
\beq{\cal L}_{AVV}=g\tr\Big[\Big([V_\nu,\pd_+V_\nu]-
[\pd_{\nu} V_\nu,V_+]-2[V_\nu,\pd_{\nu} V_+]-V_+\pd_+^{-1}V_+\pd_\perp^2\Big)
A_-\Big].
 \label{e14}\eeq
The three first terms give the standard vertex
\beq
-\frac{g}{2}f^{b_1cd}\Big( (k+p)_+g_{\mu\sigma}+(p-2k)_\mu n^+_\sigma
+(k-2p)_\sigma n^+_\mu\Big)
\label{e15}\eeq
and the last term defines the induced vertex
\beq
gf^{b_1cd}\frac{q_1^2}{2p_+}n_\mu^+n_\sigma^+.
\label{e16}\eeq

\section{Calculation of diagrams}
The total contribution to the vertex reggeon $\to$ two reggeons $+$ particle
is represented by four diagrams shown in Fig. 2, in which particles are shown by
solid lines and reggeons by wavy lines.
\begin{figure}[ht]
\epsfxsize 4in
\centerline{\epsfbox{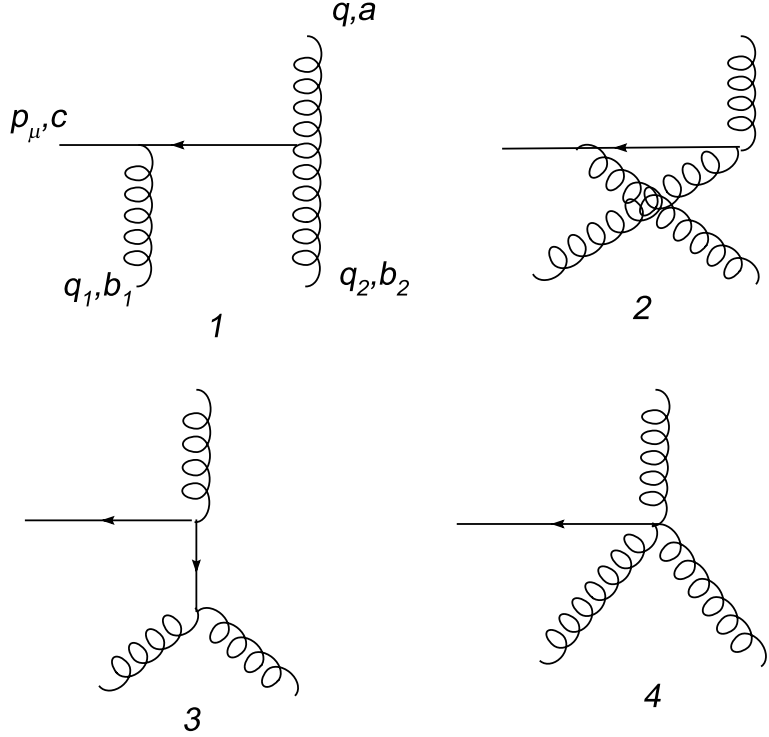}}
\caption{Reggeon diagrams for the vertex reggeon$\to$ 2reggeons
+particle}
\label{Fig2}
\end{figure}

\subsection{Diagram Fig. 2.3}
We start from the simplest diagram Fig. 2.3.
For this diagram  we have to know the vertex $<VA_-A_->$.
The Lagrangian term with two fields $A_-$ is given by \Ref{e10}.
The standard triple vertex is obtained as a product of the first and
second terms in the brackets after taking the square. However it vanishes,
since the first term in the brackets is zero.
It is important here that expansion of the $P$-exponential in the
effective action formalism generates only  an induced vertex
$\langle VA_-A_+\rangle$ but not  $\langle VA_-A_-\rangle$

Thus diagram Fig. 2.3 does not give any contribution in our kinematics.

\subsection{Diagram Fig. 2.4}

For the diagram Fig. 2.4 we have to know $<A_+A_-A_-V_\mu>$. It is contained in
the
part of the Lagrangian \Ref{e12}.
In it factor $1/D_-$ is a remnant of the $P$-exponential:
\beq
U(V_-)=P\exp\Big(-\frac{g}{2}\int_{-\infty}^{x_+}dx'_+V_-(x'_+)\Big),
\label{e17}\eeq
so that
\beq
\pd_-U(V_-)=2\frac{\pd}{\pd x_+}U(V_-)=-gV_-U(V_-).
\label{e18}\eeq
Taking into account that $U_{g=0}=1$ we get
\beq
U(V_-)=\sum_{n=0}(-g\pd_-^{-1}V_-)^n=\frac{1}{D_-}\pd_-.
\label{e19}\eeq

The terms of  ${\cal L}_{AAAV}$ which determine the vertex with only one
gluon are
\[ -\frac{g}{2}\tr\Big(-g[V_+,A_-][A_-,A_+]-2g(V_-\frac{1}{\pd_-}A_-
\frac{1}{\pd_-}A_-\pd_{\perp}^2A_+
\]\beq
+A_-\frac{1}{\pd_-}V_-\frac{1}{\pd_-}A_-
\pd_{\perp}^2A_++ A_-\frac{1}{\pd_-}A_-\frac{1}{\pd_-}V_-\pd_{\perp}^2A_+)
\Big).
\label{e20}\eeq
The first term is the standard quartic interaction, others give the
induced interaction.

We pass to the momentum representation: for the incoming momentum
$i\pd_\mu\to p_\mu$ for the outgoing $i\pd_\mu\to -p_\mu$. We find
accordingly
\beq
 \frac{1}{\pd_-}A_-(q_1)\to \frac{i}{-q_{1-}},\ \
\frac{1}{\pd_-}V_-(p)\to \frac{i}{-p_{-}},\ \
\pd^2A_+(q)\to-q_\perp^2\simeq-q^2.
\label{e21}\eeq
The last identity follows from $\pd^2A_+=(\pd_+\pd_-+\pd_{\perp}^2)A_+=
\pd^2_\perp A_+$. If the derivative acts on the product of fields:
$\pd_-(A_1A_2...A_n)$ then it corresponds to the sum of momenta in the
momentum representation. It is also true for the inverse operator
$1/\pd_-$: it corresponds to the sum of momenta in the denominator.

Note the kinematical relations among the momenta:
\[ p=q-q_1-q_2,\ \ q_-<<p_-\sim q_{1-}\sim q_{2-},\ \
q_{1+}\sim q_{2+}<<p_+=q_+,\ \ p_-=-q_{1-}-q_{2-},
\]\beq
q_\perp\sim q_{1\perp}\sim q_{2\perp}\sim p_{\perp}.
\label{e22}\eeq
Each vertex bears factor $i$, the diagram as a whole has factor $1/i$ for
the amplitude, which is omitted.

 We find for the induced vertex the following. The common factor is
 $g^2i(-i)^4(-i)^2 n_{\mu}^-$ (where $(-i)$ from each of the fields,
 $(-i)^2$ from two $1/\pd_-$, $n_{\mu}^-$  from $V_-$). We use $q_-=0$,
 so that $1/\pd_-$ does not act on $A_+$).
 \[ V_4^{ind}=
 -ig^2n_{\mu}^-(-q_\perp^2)\tr\Big(
 \frac{T^cT^{b_1}T^{b_2}T^a}{(q_{1-}+q_{2-})q_{2-}}+
\frac{T^{b_1}T^cT^{b_2}T^a}{(p_-+q_{2-})q_{2-}}+
\frac{T^{b_1}T^{b_2}T^cT^a}{(p_-+q_{2-})p_-}\Big)+\Big(q_1\leftrightarrow
q_2,\ b_1\leftrightarrow b_2\Big)\]\[=
 -ig^2n_\mu^-q_\perp^2\tr\Big(
 \frac{1}{p_-q_{2-}}(T^cT^{b_1}T^{b_2}T^a+T^{b_2}T^{b_1}T^cT^a)\]\beq+
 \frac{1}{p_-q_{1-}}(T^{b_1}T^{b_2}T^cT^a+T^cT^{b_2}T^{b_1}T^a)+
 \frac{1}{q_{1-}q_{2-}}(T^{b_1}T^cT^{b_2}T^a+T^{b_2}T^cT^{b_1}T^a)\Big).
\label{e23}\eeq
We have used: $q_{1-}+q_{2-}=-p_-$.
Also using
\[
\frac{1}{p_-q_{1-}}+\frac{1}{p_-q_{2-}}=-\frac{1}{q_{1-}q_{2-}}\]
and the cyclic symmetry under the trace we find
\[
V_4^{ind}=
 -ig^2n_\mu^-q_\perp^2\tr\Big(
 \frac{1}{p_-q_{2-}}([T^cT^{b_1}]T^{b_2}T^a+T^{b_2}[T^{b_1}T^c]T^a)\]\[+
 \frac{1}{p_-q_{1-}}(T^{b_1}[T^{b_2}T^c]T^a+[T^cT^{b_2}]T^{b_1}T^a)\Big)
 \]\beq=
 i\frac{g^2}{2}n_\mu^-q_\perp^2\Big(
 \frac{1}{p_-q_{2-}}f^{b_1cd}f^{ab_2d}+\frac{1}{p_-q_{1-}}f^{ab_1d}f^{b_2cd}
 \Big).
 \label{e24}\eeq

To find the contribution from the quartic vertex we have only to introduce
the indices $\pm$ into the general expresion and use the
light-cone values of $g_{\mu\nu}$. We get (same as from the effective action)
\beq
V_4=-i\frac{g^2}{4}n_\mu^+(f^{b_1cd}f^{ab_2d}+f^{ab_1d}f^{b_2cd}).
\label{e25}\eeq

So the total contribution from the diagram Fig. 2.4 is
\beq
V_4^{tot}=\frac{ig^2}{4}\Big[
f^{b_1cd}f^{ab_2d}\Big(2\frac{q_\perp^2n_\mu^-}{p_-q_{2-}}-n_\mu^+\Big)+
f^{ab_1d}f^{b_2cd}\Big(2\frac{q_\perp^2n_\mu^-}{p_-q_{1-}}-n_\mu^+\Big)
\Big].
\label{e26}\eeq
In the gauge $V_+=0$
convolution of  \Ref{e26}  with  polarization vector \Ref{e8}  gives
\beq
ig^2\frac{p\epsilon_\perp}{p^2_\perp}q_{\perp}^2
\Big(\frac{1}{q_{2-}}
f^{b_1cd}f^{ab_2d}+\frac{1}{q_{1-}}f^{b_1ad}f^{cb_2d}\Big)\label{e27}\eeq
Note that only the induced vertex contribution remains, the contribution
from the standard quartic vertex vanishes.

\section{Diagrams Figs. 2.1 and 2.2}
We pass to diagrams shown in Figs. 2.1 and 2.2. They only differ by
the permutation of reggeons
$A_-$, that is $q_1\leftrightarrow q_2,\  b_1\leftrightarrow b_2$. So we
study only the first one. To write the expresion we have to know vertexes
reggeon $\to$ reggeon+ gluon and reggeon-gluon-gluon.

The first (Lipatov) vertex is known and given by \Ref{e7}.
It is instructive to see how it is derived from \Ref{e11}.
In the momentum representation $A_+\to (-iq,a)$,
$A_-\to (iq_2,b)$ and $V\to (ik,d,\sigma)$. Factors: $(-i)^3$ from the
fields, $(-i)$ from derivatives and $i/2$ from the commutator and trace.
We get for the standard vertex:
\[
\frac{(-i)^4igi}{2}\Big( -q_{2\sigma}^\perp f^{dab_2}+
q_{2\sigma}^\perp f^{db_2a}+\frac{1}{2}n^-_\sigma(-k_+)f^{b_2ad}+
\frac{1}{2}n^+_\sigma(-k_-)f^{ab_2d}\Big)\]\[=
\frac{g}{2}f^{ab_2d}\Big(q_{2\sigma}^{\perp}+q_\sigma^{\perp}-
\frac{1}{2}n^-_\sigma k_++\frac{1}{2}n^+_\sigma k_-\Big)\]\beq=
\frac{g}{2}f^{ab_2d}\Big(q_{2\sigma}^{\perp}+q_\sigma^{\perp}-
\frac{1}{2}n^-_\sigma q_+-\frac{1}{2}n^+_\sigma q_{2-}\Big)=
\frac{g}{2}f^{ab_2d}\Big((q+q_2)_\sigma
-n_\sigma^-q_+-n^+_\sigma q_{2-}\Big).
\label{e28}\eeq
This coincides with the expression for the standard triple vertex
$gf^{db_2a}\gamma_{\sigma+-}(k,q_2)$.
Now the induced vertex ($\pd_+^{-1}A_+\to i/q_+$,
$\pd_-^{-1}A_-\to -i/q_{2-}$)
\beq
-\frac{1}{2}(-i)^4igi\Big(-\frac{1}{q_+}n^+_\sigma f^{dab_2}(-q_{2\perp}^2)+
\frac{1}{q_{2-}}n^-_\sigma f^{db_2a}(-q_{2\perp}^2)\Big)=
\frac{1}{2}gf^{ab_2d}\Big(\frac{q_2^2}{q_+}n^+_\sigma+\frac{q^2}{q_{2-}}
n^-_\sigma\Big).
\label{e29}\eeq
The full vertex is then given by \Ref{e7}.
It has to be transversal. Indeed convolution with $k_\sigma=(q-q_2)_\sigma$
gives
\beq
(q-q_2)_\sigma (q+q_2)_\sigma+\Big(\frac{q_2^2}{q_+}-q_{2-}\Big)q_++
\Big(\frac{q^2}{q_{2-}}-q_+\Big)(-q_{2-})=q^2-q_2^2+q_2^2-q^2-
q_{2-}q_++q_+q_{2-}=0\label{e30}\eeq
Further simplification is possible in the gauge $V_+=0$: then terms with
$n^+_{\mu}$ go and
the vertex acquires the form \Ref{e9}.

To calculate diagram Fig. 2.1 we finally need the vertex gluon$\to$reggeon+gluon
(Fig. 1.2).
The relevant terms in the effective action are \Ref{e14}.
Using the kinematical relation $k=p+q_1$, which leads to $p-2k=-p-2q_1$,
$k-2p=q_1-p$ and $k_+=p_+$, we find from \Ref{e15} and \Ref{e16}
the total vertex as
\beq
\frac{gf^{b_1cd}}{2}\Big(-2p_+g_{\mu\sigma}+(p+2q_1)_\mu n^+_\sigma+
(p-q_1)_\sigma n^+_\mu +\frac{q_1^2}{p_+}n^+_\mu n^+_\sigma\Big).
\label{e31}\eeq

Diagram Fig. 2.1 is obtained by convoluting this vertex with the vertex $AAV$
and the gluon propagator $-ig_{\sigma\sigma'}\delta_{dd'}/k^2$.
We find
\[
\frac{-ig^2f^{b_1cd}f^{ab_2d}}{4k^2}
\Big(-2p_+g_{\mu\sigma}+(p+2q_1)_\mu n^+_\sigma+
(p-q_1)_\sigma n^+_\mu +\frac{q_1^2}{p_+}n^+_\mu n^+_\sigma\Big)
\]\[
\Big[ q_\sigma+q_{2\sigma}+
\Big(\frac{q_2^2}{q_+}-q_{2-}\Big)n^+_\sigma+
\Big(\frac{q^2}{q_{2-}}-q_+\Big)n^-_\sigma\Big]
\]\[
=
\frac{-ig^2f^{b_1cd}f^{ab_2d}}{4k^2}
\Big[ q_+(-2q-2q_2+p+2q_1)_\mu+((q+q_2)(p-q_1)+q_1^2)n^+_\mu+
\Big(\frac{q_2^2}{q_+}-q_{2-}\Big)(-q_+)n^+_\mu+
\]\beq
\Big(\frac{q^2}{q_{2-}}-q_+\Big)\Big(-2q_+n^-_\mu+2(p+2q_1)_\mu+
(p_--q_{1-}+2\frac{q_1^2}{q_+})n^+_\mu\Big)\Big].
\label{e32}\eeq

Diagram Fig. 2.2. is obtained by the interchange of reggeons 1 and 2.

\subsection{Total vertex}

The total vertex is given by the sum of diagrams corresponding to
Figs. 2.1, 2.2 and 2.4 (since the contribution of Fig. 2.3 is zero)
The sum of diagrams Fig. 2. 1+2+4 is found to be transversal (convolution with
$p_\mu$ is zero) as it should be.

To find the transition amplitude we have to convolute the sum of diagrams
Fig. 2.1+2+4 with the polarization vector $\epsilon^*_\mu(p)$ and divide by $i$.
In the gauge $V_+=0$ we have
\beq
 \epsilon_\mu(p)=
\epsilon^{\perp}_\mu-\frac{(p\epsilon_\perp)}{p^+}n^+_\mu.
\label{e33}\eeq
It satisfies
\beq\epsilon p=0,\ \ \epsilon_+=0,\ \ \epsilon_-=-\frac{2(p\epsilon_\perp)}{p_+}
\label{e34}\eeq
on the mass shell $p^2=0$.
The result of the convolution is
\[
g^2\frac{f^{b_1cd}f^{ab_2d}}{(q-q_2)^2}\Big[q_+(q\ep)+
\frac{q^2}{q_{2-}}
\Big(-(q-q_2)\ep+\frac{(q-q_2)^2}{p_\perp^2}(p\ep)\Big)\Big]
\]\beq
+g^2\frac{f^{b_2cd}f^{ab_1d}}{(q-q_1)^2}\Big[q_+(q\ep)+
\frac{q^2}{q_{1-}}
\Big(-(q-q_1)\ep+\frac{(q-q_1)^2}{p_\perp^2}(p\ep)\Big)\Big]
\label{e35}\eeq
If we denote:
\beq
 u_1=q-q_1,\ \ u_2=q-q_2,\ \ F_1=f^{b_2cd}f^{ab_1d},\ \
F_2=f^{b_1cd}f^{ab_2d}
\label{e36}\eeq
Then \Ref{e35} can be rewritten as
\beq
g^2F_1\Big(\frac{q_+}{u_1^2}(q\ep)+\frac{q^2}{u_1^2u_{1-}}(u_1\ep)-
\frac{q^2}{u_{1-}p_\perp^2}(p\ep)\Big)+
g^2F_2\Big(\frac{q_+}{u_2^2}(q\ep)+\frac{q^2}{u_2^2u_{2-}}(u_2\ep)-
\frac{q^2}{u_{2-}p_\perp^2}(p\ep)\Big)
\label{e37}\eeq

\section{The diffractive amplitude}
The obtained R$\to$RRP vertex (35) has a 4-dimensional form. To reduce it to the
transverse vertex we have to study a concrete amplitude which involves this vertex.
We choose the simplest amplitude possible: production of a real gluon in collision of
two quarks, the target quark interacting with the two final reggeized gluons
in the colourless state (Fig. 3).
\begin{figure}[ht]
\epsfxsize 4in
\centerline{\epsfbox{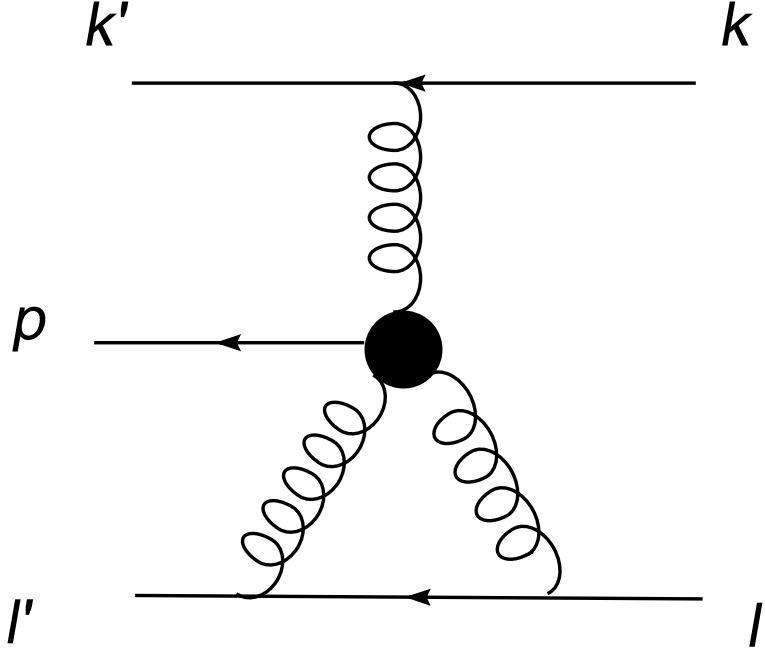}}
\caption{The difractive diagram with a R$\to$RPP vertex}
\label{Fig3}
\end{figure}
This is a simplified part of the diffractive process studied
in ~\cite{BW}, where the projectile was taken to be a quark loop. Note that
there are additional diagrams for this process
which involve two reggeon exchange between
projectile and target with the gluon emitted by the Lipatov vertex
(Fig.4). However as we shall presently see (and as is known) for them
integration over longitudinal variables in the loop presents no difficulties,
so that all the problem is concentrated in the diagram with the found
R$\to$RRP vertex.
\begin{figure}[ht]
\epsfxsize 4in
\centerline{\epsfbox{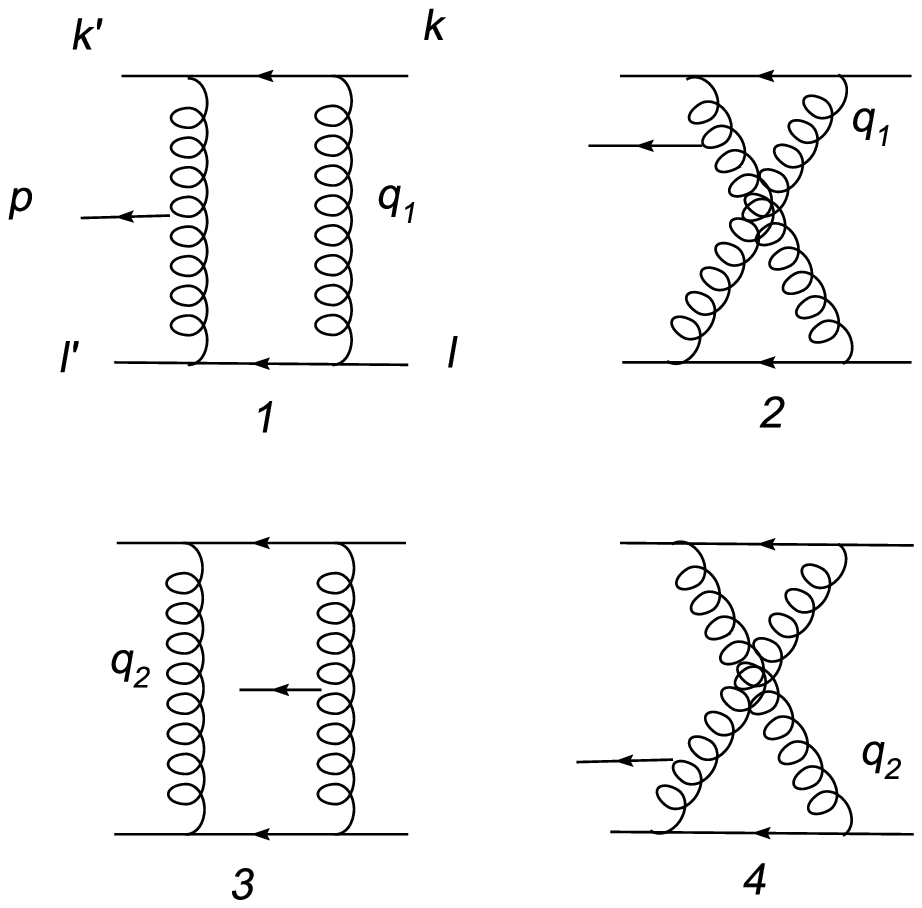}}
\caption{Diffractive diagrams with two-reggeized-gluon exchange}
\label{Fig4}
\end{figure}

\subsection{Two-reggeon exchange}
To write the expression for the contribution of two-reggeon exchange
(Fig. 4) we have to know the impact factors. For colliding quarks
they are trivial. Let the projectile and target initial momenta
be $k$ and $l$ respectively, with $k_-=l_+=k_\perp=l_\perp=0$.
Then, say, for the projectile we find the part of the diagram
coupled to the reggeons as
\beq
 \frac{1}{(k-q_1)^2+i0}
u't^{a_2}\gamma_{\mu_2}(\hat{k}-\hat{q}_{1})\gamma_{\mu_1}t^{a_1}u.
\label{e38}\eeq
In the BFKL kinematics, as is well-known, we can substitute
\footnote{In this section we use the normalization
$ab=a_+b_-+a_-b_++(ab)_\perp$}
\beq
\gamma_{\mu}\to\frac{\hat{l}}{l_-}
\label{e39}\eeq
where $l$ is the initial momentum of the target. The spin
matrix element which results is
\beq u'\hat{l}\hat{k}\hat{l}u=4(kl)^2\label{e40}\eeq
where we neglected $q_{1+}$ and $q_{2+}$ as compared with
$k_+$ in our kinematics. With a similar calculation for the target
we find the overall  factor coupled to the vertex in Fig. 4.1  as
\beq
\frac{16(kl)^2}{(k-q_1)^2(l+q_1)^2}u't^{a_2}t^{a_1}u\cdot
w't^{b_2}t^{b_1}w
\label{e41}\eeq
where $u$, $u'$  and $w$, $w'$  are the initial and
final colour wave functions of the projectile and target.
In the crossed diagram Fig. 4.2 the projectile plus target factor is
\beq
\frac{16(kl)^2}{(k-q_1)^2(l'-q_1)^2}u't^{a_2}t^{a_1}u\cdot
w't^{b_1}t^{b_2}w
\label{e42}\eeq
The impact factors for diagrams of Fig.4.3 and Fig 4.4 have a
similar structure with the substitution $q_1\to -q_2$ and
$k,\leftrightarrow k',l'$

The overall colour factor is
obtained by the convolution of colour factors from impact factors and
the vertex. We have also to take into account the projector onto the
colourless state of the reggeons coupled to the target, which is
proportional to $\delta_{b_1b_2}$
Combining these factors we find the colour operator acting on the
projectile quark for the diagrams Fig. 4.1 and 4.2 as
\beq Q^c=f^{a_2ca_1}t^{a_2}t^{a_1}\label{e43}\eeq
and $-Q^c$ for the diagrams Fig.4.3 and Fig. 4.4.
The colour operator acting on the target quark is naturally a unit
operator.

Now we pass to the longitudinal integrations in the diagrams of Fig. 4.
We start from the diagram Fig. 4.1 and integrate over $q_{1+}$.
The two denominators  are
\beq [2q_{1-}(q_{1+}-k_+)+q_{1\perp}^2+i0][2q_{1+}(l_-+q_{1-})+q_{1\perp}^2+i0].
\label{e44}\eeq
A non-zero result is obtained only when the two poles in $q_{1+}$ are on the opposite
sides from the real axis. This limits the integration over $q_{1-}$:
\beq
 -l_-<q_{1-}<0.
\label{e45}\eeq
Denoting the expression for the part of the diagram depending on the
longitudinal momenta (the inverse of \Ref{e44}) as $D_1$ and taking the residue we
find
\beq \int\frac{dq_{1+}}{2\pi i}D_1=
\frac{1}{4k_+q_{1-}^2+4k_+l_-q_{1-}-2l_-q_{1\perp}^2}.
\label{e46}\eeq
Subsiquent integration in $q_{1-}$ in the interval $-l_-<q_{1-}<0$
gives
\beq
\int_{-l_-}^0dq_{1-}\int\frac{dq_{1+}}{2\pi i}D_1=
\frac{1}{2(kl)}\Big(\ln\frac{|q_{1\perp}^2|}{2(kl)}+\pi i\Big).
\label{e47}\eeq
Of course the imaginary part could be immediately found by Cutkoski rules.

In the crossed diagram Fig. 4.2 the second denominator in \Ref{e44} is
changed to
\beq
 -2(q_{1+}-l'_+)(l'_--q_{1-})+(l'-q_1)_{\perp}^2+i0.
\label{e48}\eeq
Now a non-zero result is obtained for $0<q_{1-}<l'_-\simeq l_-$.
Taking the residue in $q_{1+}$ at the same pole as before we find
\beq\int\frac{dq_{1+}}{2\pi i}D_2=
-\frac{1}{4k_+q_{1-}^2-4k_+l_-q_{1-}+2l_-q_{1\perp}^2}
\label{e49}\eeq
and the subsequent integration over $q_{1-}$ gives
\beq
\int_{0}^{l_-}dq_{1-}\int\frac{dq_{1+}}{2\pi i}D_2=
-\frac{1}{2(kl)}\ln\frac{|q_{1\perp}^2|}{2(kl)}.
\label{e50}\eeq
Naturally the crossed diagram contains no imaginary part.
In the sum the real parts cancel and the final result is
\beq
\int\frac{dq_{1+}dq_{1-}}{2\pi i}(D_1+D_2)=
\frac{\pi i}{2(kl)}
\label{e51}\eeq
Together with the color factor and factors depending on the transverse
momenta this gives the amplitude
\beq
f^{a_2ca_1}u't^{a_2}t^{a_1}u\cdot 4(kl)\frac{1}{q_{1\perp}^2
q_{2\perp}^2}\Big(\frac{(q\epsilon)_\perp}{q_\perp^2}-
\frac{(p\epsilon)_\perp}{p_{\perp}^2}\Big),\ \ q_2=q-p-q_1.
\label{e52}\eeq
This is the standard form for the amplitude described by
Figs 4.1 and 2 in  the transverse momentum space
(see e.g. ~\cite{BW}).

Note that the same result is obtained if one passes from variables
$q_{1+}$ and $q_{2-}$ to energies squared $s_1=(k-q_1)^2$ and
$s_2=(l+q_1)^2$ for the upper and lower quark-reggeon amplitudes
in Fig. 4.1 and  deforms the Feynman contours of integration to
pass around the unitarity cuts at $s_{1,2}\geq 0$. The contribution from
the intermediate quark states at $s_1=s_2=0$ will give directly (50).

Diagrams of Fig. 4.3 and Fig. 4.4 are calculated in the same manner and
after integration over the longitudinal variables reduce to the
standard formulas in terms of the transverse variables.

\subsection{ Contribution from the R$\to$ RRP vertex}
The two terms in the vertex \Ref{e35} which differ by the permutation
$1\leftrightarrow 2$ evidently give the same contribution after
integration over the loop momentum and summation over colour indices.
So it is sufficient to study only one of them. To use the previous results
we choose the second term.
The colour operator applied to the projectile quark is given by
\beq
f^{b_2cd}f^{ab_1d}\delta_{b_1b_2}t^a=-N_ct^a\delta_{ac}\label{e53}\eeq
and is common for the three terms in the second part of \Ref{e35}.

The first term in this second part, proportional to $(q\epsilon)_\perp$ has the
denominators
in the direct and crossed terms which are obtained from \Ref{e44} and \Ref{e48}
by changing $k\to q$. So the first denominator becomes
\beq 2q_{1-}(q_{1+}-q_+)+(q-q_1)_{\perp}^2+i0
\label{e54}\eeq
Obviously after integration we shall get the same formulas \Ref{e47} and
\Ref{e50}
in which we have to substitute (ql) instead of (kl) and $(q-q_1)_\perp^2$
instead of
$q_{1\perp}^2$.
The real parts again cancel and the result will be given by
\beq
\int\frac{dq_{1+}dq_{1-}}{2\pi i}(D_1+D_2))=
\frac{\pi i}{2(ql)}
\label{e55}\eeq
where $D$'s  refer only to the parts coming from the two propagators
in the direct and crossed terms.
This has to be multiplied by the colour factor and momentum factors coming
from the two impact factors. The latter for the target is given by an expresion
similar to \Ref{e38}
and for the projectile is just $2k_+$. Taking into account the
rest of the factors   we finally have the complete contribution from the first
term in the second part of \Ref{e35} as
\beq
\int\frac{dq_{1+}dq_{1-}}{2\pi i}T_1=
4\pi i(pl)\delta_{ac}u't^au\cdot
\frac{1}{q_{1\perp}^2q_{2\perp}^2}\frac{q\epsilon_\perp}{q_{\perp}^2}
\label{e56}\eeq
This coincides with the first term in the contribution from the Bartels
vertex to the diffractive amplitude.

As in the previous calculation of the two-reggeon exchange the same
result is obtained if one similarly introduces energies squared
$s_1=(q-q_1)^2$ and
$s_2=(l+q_1)^2$ for the amplitudes  R+R$\to$ P+R and q+R$\to$ q+R
and then deforms the Feynman integration contour to close on the
unitarity cuts of both amplitudes. The contribution from
the intermediate  quark+gluon state will immediately give (55).

The situation with the second term is a bit more complicated.
In both the direct and crossed term an extra factor $1/q_{1-}$ appears.
The integration over $q_{1+}$ is done exactly as before. However the
subsequent integration over $q_{1-}$ cannot be done separately for the
direct and crossed term
because of the singularity at $q_{1-}=0$. However in their sum this singularity
cancels and integration becomes trivial. In fact after the integration
over $q_{1+}$ we find the integral over $q_{1-}$:
\[ \int_{-l_-}^0\frac{dq_{1-}}{q_{1-}}\int\frac{dq_{1+}}{2\pi i}D_1=
\int_{-l_-}^0\frac{dq_{1-}}{q_{1-}} \
\frac{1}{4k_+q_{1-}^2+4k_+l_-q_{1-}-2l_-(q-q_1)_{\perp}^2}\]\beq=
-\int_{0}^{l_-}\frac{dq_{1-}}{q_{1-}} \
\frac{1}{4k_+q_{1-}^2-4k_+l_-q_{1-}-2l_-(q-q_1)_\perp^2},
\label{e57}
\eeq
whereas the corresponding integral of the crossed
term is
\beq
\int_{0}^{l_-}\frac{dq_{1-}}{q_{1-}}\int\frac{dq_{1+}}{2\pi i}D_2=
-\int_{0}^{l_-}\frac{dq_{1-}}{q_{1-}} \
\frac{1}{4k_+q_{1-}^2-4k_+l_-q_{1-}+2l_-(q-q_1)_\perp^2}
\label{e58}
\eeq
The sum of \Ref{e57} and \Ref{e58} is regular at
$q_{1-}=0$. So integration of this sum can be done directly. The
resulting real part again vanishes and one finds
\beq
\int\frac{dq_{1+}dq_{1-}}{2\pi i q_{1-}}(D_1+D_2))= -\frac{\pi
i}{2l_-(q-q_1)^2_\perp} \label{e59}
\eeq
Adding all the rest
factors we get for the second term in the second part of \Ref{e35}
\beq
\int\frac{dq_{1+}dq_{1-}}{2\pi i}T_2= -4\pi
i(pl)\delta_{ac}u't^au\cdot
\frac{1}{q_{1\perp}^2q_{2\perp}^2}\frac{(q-q_1)\epsilon_\perp}
{(q-q_1)_{\perp}^2} \label{e60}
\eeq
This gives the second part of
the contribution corresponding to the Bartels vertex.

Note that if one tries to use here the same simplified method of
integration over energies of the R+R$\to$R+P and q+R$\to$ q+P
amplitudes closing the contour arond the unitarity cuts, then
one encounters the singularity at $q_{1-}=0$ with an unknown way of
integration around it. If one just neglects this singularity, that is
takes into account
only the unitarity contribution to the dicontinuities of the
amplitudes, then one gets a result which is
twice larger than (59) and so incorrect.

We are left with the third term in the second part of \Ref{e35}
 with a structure which has no
counterpart in the Bartels vertex. For the 4 dimensional and 2-dimensional
pictures to coincide this contribution has to disappear.

The only denominator in the direct term is
\beq
 (l+q_1)^2+i0=2q_{1+}(l_-+q_{1_-})+q_{1\perp}^2+i0.
\label{e61}\eeq
Obviously the integral over $q_{1+}$ is divergent and can be
studied only for the sum of this direct and crossed term. For the latter the
denominator is
\beq
 (l'-q_1)^2+i0=-2(q_{1+}-l'_+)(l'_{-}-q_{1-})+(l'-q_1)_{\perp}^2+i0.
\label{e62}\eeq
In the integrand at large $q_{1+}$ we find a sum
\beq
 \frac{1}{2q_{1+}(l_-+q_{1_-})}-\frac{1}{2q_{1+}(l'_--q_{1_-})}
\label{e63}\eeq
Since $l'=l+q-p$ we have $l'_--q_{1-}=l_-+q_--p_--q_{1-}
\simeq l_--p_--q_{1-}$ and is generally not equal to $l_-+q_{1-}$.
Therfore at $q_{1+}\to\infty$  the sum does not generally go to zero
faster than $1/q_{1+}$, so that integration over $q_{1+}$ remains
divergent even after summing the direct and crossed terms.

The only possibility to give some sense to this integration is to assume
that in the target denominator one may negelect minus components of all
momenta except for the incoming target momentum $l_-$. This means that
in \Ref{e61} and \Ref{e62} we take  $l'_-=l_-$ and $q_{1-}=0$.
Then integration over
$q_{1+}$ becomes convergent and since the poles of the two terms lie
at opposite sides from the real axis, gives a non-zero result, which
does not depend on the value of $q_{1-}$. After that we find the
integral over $q_{1-}$ of the form
\beq
\int_{-\infty}^{\infty}\frac{dq_{1-}}{q_{1-}}
\label{e64}\eeq
The only possibility to give some sense to it is to assume
that it should be taken as a principal value integral. Then it is equal
to zero and the third term in \Ref{e35} disappears indeed.

As a result we find that the longitudinal integration of the 4-dimensional
vertex found by the effective action technique requires certain caution.
One finds a piece, for which a strict integration is divergent. To overcome
this diffculty one has to neglect the small minus components in the target
impact factor (and then do the $q_{1+}$ integration in the trival manner
closing the contour around the unitarity cut of the reggeon-target
amplitude) and afterwards do the remaining $q_{1-}$ integration by the
principal value recipe. This result has been found only for the
diffractive amplitude. It remains an open question if it has a wider
validity and applies also to other cases, which correspond to
double and single cuts of the general triple pomeron amplitude.

\section{Conclusions}
The Lipatov effective action has demonstrated its advantage
for the construction of vertexes for production of more than one particle
from the reggeized gluon, which are necessary for the calculation of
next-to-leading corrections to the BFKL pomeron ~\cite{nll}.
We have applied the rules of the Lipatov effective action to construct
the vertex in which a reggeon passes into two reggeons emitting a gluon.
This vertex allows to study processes in which the number of reggeized
gluons changes. As expected, the effective action allows to build
this vertex almost automatically, in contrast to earlier derivations,
in which this required a considerable effort. However the obtained
vertex is found in the 4-dimensional form and formally contains
singularities
in the longitudinal variables.

Its reduction to the pure transverse form meets with some difficulties,
related to these singularities. In the diagram with an intermediate gluon
this singularity requires to do the longitudinal integration
preserving the dependence of the target impact factor on small
longitudinal momenta. As a result the mentioned singularity disappears.
In the diagram with pointlike vertex,
on the contrary, small longitudinal momenta are to be neglected in the
target impact factor. Then the remaining singularity has to be
integrated over by the principal value recipe. After that one finally
obtains the same transverse vertex, as found earlier by direct methods.

The fact that one has to apply different rules to treat different
contributions certainly leaves a certain  feeling of uneasiness. We hope
that the only contributions which require  special treatment are those
with pointlike emission, which seem to have been dropped from the start in
previous derivations using either dispersion approach or the dominance
of large nuclear distances. If this is so then the use of the
effective action preserves its unversality except for these
exceptional cases. To verify this one has to study more complicated
processes  involving the found vertex, such as  double scattering off
a colorless target. This study is now in progress.

\section{Acknowledgements}
The authors have benefited from constructive discussions with
J.Bartels, L.Lipatov and G.P.Vacca, whose comments they highly
appreciate. M.A.B. is thankful to Hamburg University for hospitality and
financial support. This work has also been supported by grants
RNP 2.1.1.1112 and RFFI 06-02-16115a of Russia.

\end{document}